\PassOptionsToPackage{binary-units}{siunitx}

\documentclass[conference]{IEEEtran}
\usepackage[utf8]{inputenc}

\usepackage[english]{babel}
\usepackage[english]{isodate}
\usepackage{blindtext}

\usepackage{cite}

\usepackage[acronym,smallcaps]{glossaries}
\newacronym{puf}{puf}{Physical Unclonable Function}
\newacronym{ro}{ro}{ring-oscillator}
\newacronym{sram}{sram}{static random access memory}
\glsunset{sram} 
\newacronym{fpga}{fpga}{field programmable gate array}
\glsunset{fpga} 
\newacronym{asic}{asic}{application specific integrated circuit}
\glsunset{asic} 
\newacronym{kde}{kde}{kernel density estimation}
\newacronym{pca}{pca}{principal component analysis}
\newacronym{pwc}{pwc}{pairwise comparison}
\newacronym{cots}{cots}{commercial off-the-shelf}
\newacronym{uc}{mcu}{microcontroller}
\newacronym{clb}{clb}{configurable logic block}
\glsunset{clb} 
\newacronym{snd}{snd}{standard normal distribution}
\newacronym{trng}{trng}{True Random Number Generator}
\newacronym{ctw}{ctw}{Context Tree Weighting}
\newacronym{hw}{hw}{Hamming weight}
\newacronym{hd}{hd}{Hamming distance}
\newacronym{crp}{crp}{challenge-response pair}
\newacronym{lut}{lut}{look-up table}
\glsunset{lut} 
\newacronym{ip}{ip}{intellectual property}
\glsunset{ip} 
\newacronym{ci}{ci}{confidence interval}
\newacronym{pdf}{pdf}{probability density function}
\newacronym{cdf}{cdf}{cumulative distribution function}
\newacronym{iid}{iid}{independent and identically distributed}
\newacronym{rv}{rv}{random variable}
\newacronym{clt}{clt}{central limit theorem}
\newacronym{far}{far}{false acceptance rate}
\newacronym{frr}{frr}{false rejection rate}

\ifCLASSINFOpdf
\else
\fi
\usepackage{graphics,graphicx}
\usepackage{epstopdf} 
\usepackage{color}
\usepackage{tikz}
\usetikzlibrary{matrix,positioning,calc}
\graphicspath{{./figures/}}

\usepackage{amsmath}
\usepackage{amssymb,amscd,upgreek,wasysym,mathtools}
\usepackage[alsoload=binary,obeyall,per=fraction]{siunitx}

\newcommand\numberthis{\addtocounter{equation}{1}\tag{\theequation}}

\newcommand{\normalQ}{\mathcal{N}^{-1}} 
\newcommand{\binomial}{\mathcal{B}\mathrm{in}} 
\newcommand{\betaQD}{\mathcal{B}\mathrm{eta}^{-1}} 
\newcommand{\bernoulli}{\mathcal{B}\mathrm{ern}}
\newcommand{\Hnull}{\mathrm{H}_\mathrm{0}} 
\newcommand{\Halt}{\mathrm{H}_\mathrm{A}} 

\DeclareMathOperator{\prob}{P}

\usepackage{url}

\begin{document}
%
\title{On the Confidence in Bit-Alias Measurement\\ of Physical Unclonable Functions}

\bstctlcite{IEEEexample:BSTcontrol}

\author{\IEEEauthorblockN{Florian Wilde and Michael Pehl}
\IEEEauthorblockA{Technical University of Munich\\
\{florian.wilde, m.pehl\}@tum.de}}


%


\maketitle

\begin{abstract}
Physical Unclonable Functions (PUFs) are modern solutions for cheap and secure key storage.
The security level strongly depends on a PUF's unpredictability, which is impaired if certain bits of the PUF response tend towards the same value on all devices.
The expectation for the probability of 1 at some position in the response, the \emph{Bit-Alias}, is a state-of-the-art metric in this regard.
However, the confidence interval of the Bit-Alias is never considered, which can lead to an overestimation of a PUF's unpredictability.
Moreover, no tool is available to verify if the Bit-Alias is within given limits.
This work adapts a method for the calculation of confidence intervals to Bit-Alias.
It further proposes a statistical hypothesis test to verify if a PUF design meets given specifications on Bit-Alias or bit-wise entropy.
Application to several published PUF designs demonstrates the methods' capabilities.
The results prove the need for a high number of samples when the unpredictability of PUFs is tested.
The proposed methods are publicly available and should improve the design and evaluation of PUFs in the future.
\end{abstract}


%
\IEEEpeerreviewmaketitle

\section{Introduction}
The need for a high level of security in low-cost devices has driven the research for cheap but secure key storage.
\Glspl{puf} are promising solutions for such applications:
They permit the derivation of a secret from chip-unique manufacturing variations, such as variations in the threshold voltage, using standard logic gates.
For this purpose, a large class of \glspl{puf} -- so called Weak \glspl{puf} or single-challenge \glspl{puf} -- consists of many measurement circuits where each evaluates its own local variations on the chip to derive one bit of the secret.
\gls{sram} \glspl{puf} \cite{Guajardo2007} and \gls{ro} \glspl{puf} \cite{Gassend2002} are just two prominent examples of such \glspl{puf}.

The security of this concept is ensured if an attacker cannot read out the \gls{puf} or predict its secret response.
The first is given if the \gls{puf} is not powered, but raises the need for dedicated runtime countermeasures while the secret is generated from the \gls{puf}.
The latter implies that an attacker who knows the output of a large number of \glspl{puf} from equally built devices, still cannot significantly reduce the entropy in the response of the \gls{puf}.
Therefore tests must be designed which ensure that there is no notable statistical weakness in the random (secret) responses of the \gls{puf}.
Multiple tests have been suggested, which measure bias \cite{Maiti2010, Hori2010}, correlations, or spatial correlations \cite{Wilde2014, Wilde2018spatcor}.
But, except for \cite{Hori2010, Wilde2018spatcor}, the accuracy of the metrics is not assessed, which can lead to undetected flaws. 

\emph{Contribution.} 
This work builds upon the Bit-Alias \cite{Maiti2010} as one well suited method to evaluate the unpredictability of a certain position in the \gls{puf} response. 
It firstly shows how to derive confidence intervals for the Bit-Alias to get the accuracy of the evaluation.
Secondly, a hypothesis test to verify if the Bit-Alias is within a selectable permissible range is introduced.
The required number of test devices to provide this guarantee with a reasonably low \gls{far} is discussed.
Thirdly, application to previous work and comparison to other metrics highlights the usefulness and limitations of our approach.

\emph{Structure.}
The rest of this work starts with 
the introduction of confidence intervals for Bit-Alias in Sec.~\ref{sec:con_test}.
Our hypothesis test is introduced in Sec.~\ref{sec:hyp_test}.
Sec.~\ref{sec:appl} and Sec.~\ref{sec:rel_to_other} demonstrate the application of the metric and discuss the results.
A conclusion is drawn in Sec.~\ref{sec:conclusion}.

\section{Statistical View on Bit-Alias\\ and Its Confidence Interval}\label{sec:con_test}
\glspl{puf} are double random:
First, manufacturing variations vary the expected behavior for each device, such as the preferred start-up value of an \gls{sram} cell or the frequency of an \gls{ro};
Second, noise and environmental effects impact the behaviour at run-time.
The former determines unpredictability, the latter reliability.
To properly investigate the unpredictability of a \gls{puf}, independent samples of solely the first random process are required.
This is commonly approximated by testing multiple devices and removing run-time randomness by averaging multiple measurements of each device.

Under the assumption that this approach produces independent samples, the Bit-Alias \cite{Maiti2010} tests whether individual positions of the \gls{puf}'s noise-free response $r$ are predictable due to an imbalance of 1s and 0s and is defined as
\begin{equation}\label{eq:ba}
  \hat{p}_t = \frac{1}{N} \sum_{n=1}^{N} r_{t,n}\text{,}
\end{equation}
where $N$ is the number of devices and $t$ indexes the bit position in a device's \SI[parse-numbers=false]{T}{\bit} response vector.
If $\hat{p}_t = 0.5$, an attacker cannot intelligently guess the response at position~$t$. 

From a statistical point of view, Eq.~\eqref{eq:ba} considers each of the $T$ bit positions as an independent Bernoulli ($\bernoulli\left(p_t\right)$) distributed \gls{rv} and estimates $p_t$, the probability for observing a 1 at position $t$, by $\hat{p}_t$, under the assumption that each device provides one independent realization of all $T$ \glspl{rv}.
Since the following applies equally to any position $t$, we omit the index from now on.

In the given scenario, the arithmetic mean~($\hat{p}$) is the best estimator for~$p$, as it is unbiased and provides uniform minimum variance \cite{Klueppelberg2018}.
Still, the estimation can be far off $p$ if it is based on few or flawed samples.
Therefore, the \gls{ci}, i.e. a range $\hat{p}_l$ to $\hat{p}_u$ that includes the true value $p$ in on average $1-\alpha$ cases, has to be taken into account.
The significance level $\alpha$ is chosen by experience, e.g. $0.05$, $0.01$.

To calculate this \gls{ci}, first note that estimating $p$ of a Bernoulli process is equivalent to estimating the proportion of the binomial distribution that results from counting the 1s in $N$ repeated trials of the process.
The binomial proportion \gls{ci} is a well explored problem and manifold methods can be found in textbooks.
We consider the following three useful for the given scenario:
First, the normal approximation interval
\begin{equation}
  \hat{p}_{l,u} = \hat{p} \pm z \sqrt{\frac{\hat{p}(1-\hat{p})}{N}}\text{,} \label{equ:wald}
\end{equation}
second, the Wilson's score interval
\begin{equation}
  \hat{p}_{l,u} = \frac{\hat{p}+\frac{z^2}{2N}}{1+\frac{z^2}{N}} \pm \frac{z}{1+\frac{z^2}{N}} \sqrt{\frac{\hat{p}(1-\hat{p})}{N}+\frac{z^2}{4\,N^2}} \label{equ:wilson}
\end{equation}
both using $\hat{p}$ and $N$ from above and with
\begin{equation}
  z = \normalQ\left(1-\frac{\alpha}{2}, 0, 1\right)\text{,}
\end{equation}
i.e. the $1-\frac{\alpha}{2}$ quantile of a standard normal distribution, and third, the so-called \emph{exact} interval by Clopper and Pearson
\begin{align*}
  \hat{p}_l &= \begin{cases}\betaQD\left(\frac{\alpha}{2},x,N-x+1\right) & x > 0\\ 0 & \text{otherwise}\end{cases} \\
  \hat{p}_u &= \begin{cases}\betaQD\left(1-\frac{\alpha}{2},x+1,N-x\right) & x < N\\ 1 & \text{otherwise}\end{cases} \numberthis\label{equ:clp}
\end{align*}
where $x$ is the number of 1s observed on $N$ devices at a given position.
The last is also used in our hypothesis test in Sec.~\ref{sec:hyp_test}.

Agresti and Coull \cite{Agresti1998} compared these and several other methods.
They found the normal approximation interval to perform poorly for small $N$, providing either too wide \glspl{ci} for $\hat{p} \approx 0.5$ or far too narrow \glspl{ci} for $\hat{p}$ close to $\{0, 1\}$.
This matches many rules of thumb which restrict the normal approximation of a binomial distribution to large $N$ with a sufficient number of both 1s and 0s.
The Clopper and Pearson method provides slightly too wide \glspl{ci}, because it ensures at least $1-\alpha$ coverage probability even at worst case values of $N$, $\hat{p}$.
In conclusion, \cite{Agresti1998} recommends Wilson's score interval, because its mean coverage probability is closest to -- though not necessarily above -- the desired level $1-\alpha$ for virtually all values of $N$, $\hat{p}$.
We consider this recommendation well suited for determining the \glspl{ci} after a \gls{puf} experiment.

\begin{figure}
\centering
\input{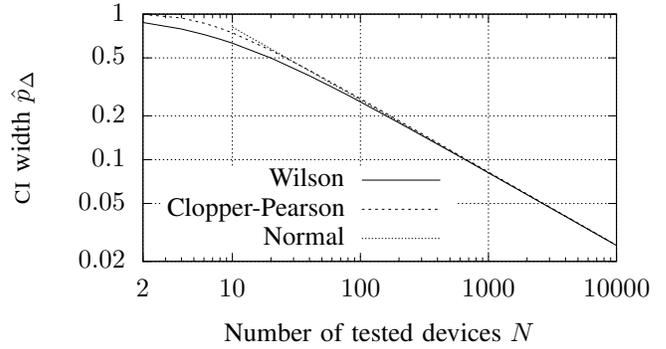}
\caption{Width of \gls{ci} according to Wilson's score, Clopper-Pearson, and normal approximation estimator over $N$ for $\alpha=0.01$, $\hat{p}=0.5$. A \gls{ci} of $0.5 \pm 0.1$ has a width of $0.2$.}\label{fig:CIwidthN}
\end{figure}

\begin{figure}
\centering
\input{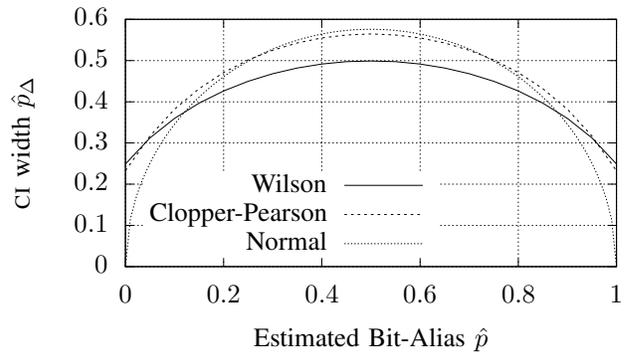}
\caption{Width of \gls{ci} according to Wilson's score, Clopper-Pearson, and normal approximation estimator over $\hat{p}$ for $\alpha=0.01$, $N=20$. A \gls{ci} of $0.5 \pm 0.1$ has a width of $0.2$.}\label{fig:CIwidthP}
\end{figure}

The normal approximation can provide -- despite its poor performance in certain cases -- a first estimate of the number of devices required for testing.
Reasonable \gls{ci} widths in real-world applications require sufficiently large $N$, cf. Fig.~\ref{fig:CIwidthN}, and $p$ is not close to $\{0, 1\}$, cf. Fig.~\ref{fig:CIwidthP}.
Given, e.g., a desired \gls{ci} width $\hat{p}_\Delta=0.1$ ($\hat{p}_{l,u} = 0.5 \pm 0.05$), and a confidence level $\alpha=0.01$ ($z=2.5759$), the number of required test devices can be roughly estimated via
\begin{equation}
  N_{\hat{p}=0.5} (\hat{p}_\Delta, z) = \left(\frac{z}{\hat{p}_\Delta}\right)^2\text{,}
\end{equation}
resulting in $664$ devices to be tested.
The Clopper-Pearson method requires $680$, the Wilson's score method $658$ devices in this setting.
Note that the provided numbers are minimum values and do not consider additional uncertainty introduced e.g. if the samples are not perfectly independent.

Independent of the chosen estimator, $\hat{p}_\Delta$ for fixed $N$ also depends on $\hat{p}$, with a maximum -- i.e. the worst case -- at $\hat{p}=0.5$, cf. Fig.~\ref{fig:CIwidthP}.
The Wilson's score interval, for example, has width $0.499$ for $\hat{p}=0.5$, but width $0.249$ for $\hat{p}=0$, when $N=20$, $\alpha=0.01$.
Thus, when comparing the number of devices for unpredictability analysis, where $\hat{p}=0.5$ is the optimum, and the number of repeated measurements required for reliability analysis, where $\hat{p}$ is close to $1$ or~$0$, the former requires twice the number of samples.

\section{Hypothesis Test on Bit-Alias\\ or Minimum Bit-wise Entropy}\label{sec:hyp_test}
\subsection{Test for Acceptable Bit-Alias} 
For the design of appropriate post-processing algorithms, a Bit-Alias within a certain range around $0.5$ must be ensured.
The following qualification test for whether the Bit-Alias of a certain position in the response vector is sufficiently close to $0.5$ is equivalent to checking whether a coin is fair.
It uses the hypothesis test that is the basis for the Clopper-Pearson \gls{ci}:

Assume as null hypothesis $\Hnull: p \geq p_u$, which we aim to reject so that the alternative hypothesis $\Halt: p < p_u$ remains.
We have to reject $\Hnull$ if the probability of observing at most $x_u$ 1s, i.e. the p-value, is too low under this hypothesis.
Due to the monotonicity of \glspl{cdf}, it is sufficient to consider the case $p = p_u$, because for $p > p_u$, the probability of observing at most $x_u$ 1s is even less.
We can therefore rule out a too high Bit-Alias value if the p-value
\begin{equation}
  p_{\mathrm{0}, u} = \prob\left[X_u \leq x_u\right] = \sum_{i=0}^{x_u} \binom{N}{i} {p_u}^i \left(1 - p_u\right)^{N-i}\label{eq:xu}
\end{equation}
with $X_u\sim\binomial\left(N, p_u\right)$ is less than $\frac{\alpha}{2}$.
The limit is $\frac{\alpha}{2}$ instead of $\alpha$, because another hypothesis test to rule out too low Bit-Alias is to be rejected simultaneously.
This second test can be constructed identically for symmetry reasons.
It has $\Hnull: p \leq p_l$, $\Halt: p > p_l$, and with $X_l\sim\binomial\left(N, p_l\right)$ p-value
\begin{equation}
  p_{\mathrm{0}, l} = \prob\left[X_l \geq x_l\right] = \sum_{i=x_l}^{N} \binom{N}{i} {p_l}^i \left(1 - p_l\right)^{N-i}\text{.}\label{eq:xl}
\end{equation}
Reversing the calculation of $\hat{p}_\Delta$ using Clopper-Pearson from Sec.~\ref{sec:con_test} provides a validation of the suggested test.
For $\alpha=0.01$, $N=680$, $p_l=0.45$, $p_u=0.55$ only $x_l=x_u=340$ should allow for the rejection of both hypotheses.
This ensures that the \gls{far}, i.e. incorrectly approving a position to be within $(p_l, p_u)$, is at most $\alpha$.

The probability to observe exactly $340$ 1s in $680$ trials, however, is even for $p=0.5$ only $0.03$.
It is therefore necessary to take the \gls{frr} into account.
Once the limits $x_u$, $x_l$ are determined from \eqref{eq:xu}, \eqref{eq:xl}, the probability to accept a position with true Bit-Alias $p$, i.e. $X\sim\binomial\left(N, p\right)$, is
\begin{equation}
  p_{\mathrm{1}} = \prob\left[x_l \leq X \leq x_u\right] = \sum_{i=x_l}^{x_u} \binom{N}{i} p^i \left(1-p\right)^{N-i}\text{.}
\end{equation}
Although $p_{\mathrm{1}}$ approaches $1$ for $N \rightarrow \infty$ if $p \in (p_l, p_u)$, for a real-world test, one would define $(p_k, p_v)$ and choose $N$ so that $\forall p \in (p_k, p_v): 1 - p_{\mathrm{1}} \leq \beta \; $.
Due to monotonicity, again a test for $\forall p \in \{p_k, p_v\} : 1 - p_{\mathrm{1}} \leq \beta$ suffices.
In the above example with $p_{l,u} = 0.5 \pm 0.05$, $\alpha=0.01$, $N=6674$ is required to achieve $\beta = 0.01$ for $p_{k,v} = 0.5 \pm 0.02$.

\subsection{Test for Early Termination of Experiment}
From a practical point of view, testing $680$ or more devices for whether too many response bits show non-satisfying Bit-Alias estimates is a waste of resources.
Therefore, a forecast whether it is reasonable to continue testing or abort the test and demand a layout change or redesign is desirable.
The previous hypothesis test can easily be adapted for this purpose.
Given, e.g., $N=50$ tested devices and some positions in the response with $x\leq10$ 1s, the hypothesis for the new test is that this low value of $x$ is by random chance and $p$ is actually at least at the lower predefined limit $p_l$, i.e. $\Hnull: p \geq p_l$.\linebreak
With $X_{l'}\sim\binomial\left(N, p_l\right)$, the p-value is
\begin{equation}
  p_{\mathrm{0}, l'} = \prob\left[X_{l'} \leq x\right] = \sum_{i=0}^{x} \binom{N}{i} {p_l}^i \left(1 - p_l\right)^{N-i}\text{.}
\end{equation}
A corresponding test for too many instead of too few 1s is easy to construct for symmetry reasons and provides another p-value $p_{\mathrm{0}, u'}$.
If $p_{\mathrm{0}, l'}$ or $p_{\mathrm{0}, u'}$ is below some $\alpha$ for too many positions, permissible Bit-Alias values are sufficiently unlikely to abort the test.
In the above example, $N=50$, $p_l=0.45$, the probability for $x \leq 10$ by random chance despite of a Bit-Alias value above $p_l$ is just $p_{\mathrm{0}, l'}=2\cdot10^{-4}$.

\subsection{Relation to Bit-Wise Entropy}
The limits $p_l$ and $p_u$ can be represented by a minimum Shannon- or min-entropy, because each position is assumed to be a Bernoulli distributed \gls{rv}, in which case entropy and probability can be directly calculated from each other.
A minimum min-entropy of $h_\infty = 0.9$, relates to $p_u = 2^{-h_\infty} \approx 0.5359$ and $p_l = 1 - 2^{-h_\infty} \approx 0.4641$.
The example from the previous section, $p = 0.5 \pm 0.05$, is equivalent to a minimum min-entropy of $0.8625$ and a Shannon-entropy of $0.9928$.
Note that the entropy values are given per position and their sum may not constitute the entropy of the entire \gls{puf} due to correlation.


\section{Application to Public PUF Data}\label{sec:appl}
\begin{figure}
\centering
\input{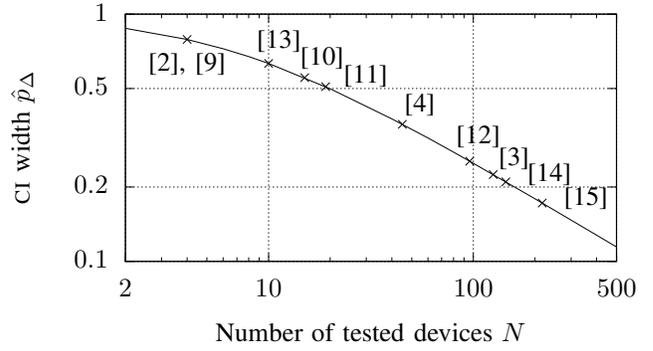}
\caption{\gls{ci} width achievable for $\alpha=0.01$, $\hat{p}=0.5$ in a collection of popular \gls{puf} publications.}\label{fig:CIwidthAnnotated}
\end{figure}

To emphasize the importance of \glspl{ci}, Fig.~\ref{fig:CIwidthAnnotated} shows the achievable $\hat{p}_\Delta$ for selected publications.
We examined almost 200 publications between 2002 and 2018 and reported those which analyzed the most devices when they were published.
Additionally some high impact papers are included for comparison.
\gls{puf} research commenced in 2002 with as few as four devices in \cite{pappu2001physical,Gassend2002} and remained with just a few devices until Maiti et al. \cite{Maiti2010} set an early record by analyzing 125 devices in 2010, increased to 193 devices in 2011.
It took until 2018 to exceed this number, by Hesselbarth et al. \cite{Hesselbarth2018}, who are the first that may claim to measure Bit-Alias with an accuracy of at least better than $\pm 0.1$.
This reveals a common issue in \gls{puf} research, where claims on superior inter-class Hamming-distance (inter-HD), Bit-Alias, or entropy are based on too few data, especially in those papers not listed in Fig.~\ref{fig:CIwidthAnnotated}, which generally analyze less than 20 devices.

\section{Relation to Other Metrics\\ and Overall \glsentrytext{puf} Tests}\label{sec:rel_to_other}
Hori et al. \cite{Hori2010} already provided \glspl{ci} for their metrics.
They utilized the fact that their metrics all calculate intermediate values per device, which are assumed to be normal distributed due to the \gls{clt}. 
Under this assumption, the \gls{ci} for the mean can then be calculated using the $t$-distribution \cite{Hori2010}, which leads to much tighter bounds than this work achieved for the Bit-Alias.
However, the approach in \cite{Hori2010} has issues.
First, the applicability of the \gls{clt} is questionable:
The \gls{clt} postulates that the arithmetic mean of a sum of \gls{iid} \glspl{rv} approaches a normal distribution.
But empirical results of strongly varying Bit-Alias \cite{Wilde2014} contradict the assumption of identical distribution, and the observation of spatial correlations \cite{Wilde2018spatcor} contradicts independence of \glspl{rv}.
Second, the metrics which Hori et al. provide a \gls{ci} for, test a mixture of several \gls{puf} properties.
This allows issues to cover each other up, e.g. too high and too low Bit-Alias at different positions of the response, and makes the interpretation of \glspl{ci} for these metrics difficult.

Beyond the approach in this paper, adapting the \gls{ci} for Bit-Alias to other metrics such as Uniformity \cite{Maiti2010} may seem tempting. 
But Uniformity measures the probability for a 1 within the response bit vector of a single device.
At the same time, the methods for \gls{ci} calculation in this work assume independent samples of the same \gls{rv}.
Consequently, the methods can only be applied to Uniformity, if all positions in the response vector of a device can be considered to be independent samples of the same \gls{rv}.
This, however, does not match the observations in previous work \cite{Wilde2018spatcor}.
For other candidate metrics, similar considerations are necessary.

Nevertheless, for a complete unpredictability evaluation of a \gls{puf} design, additional tests are required on top of our enhanced Bit-Alias.
Currently, the distribution of inter-HD and especially its arithmetic mean, named Uniqueness \cite{Maiti2010}, are usually tested.
Uniqueness, however, is entirely determined by the Bit-Alias values \cite{Pehl2016}.
Testing the Bit-Alias thus automatically ensures that the Uniqueness is within limits, and has additional benefits:
First, higher sensitivity, i.e. the numeric value of the respective Bit-Aliases changes more strongly than that of the Uniqueness, if a certain part of the response vector deviates by a certain amount from equiprobability.
Second, location information, meaning that contrary to Uniqueness, the Bit-Aliases show exactly which positions of the response bit string are biased.
This helps the designer to locate potential layout errors faster, but may also serve in a security assessment.

However, all current standard tests such as Bit-Alias, Uniqueness, or Uniformity, fail to identify correlations within the response vectors.
Therefore, a test of the inter-HD \emph{distribution}, especially its tails, or a correlation test between positions in the response vector as in \cite{Wilde2018spatcor} is mandatory.
While we consider the problem of evaluating the Bit-Alias per position solved with the additional methods presented in this work, especially the consideration of correlations between \gls{puf} response bits and the computation of confidence intervals for Uniformity or a similar metric leaves room for improvement.

\section{Conclusion}\label{sec:conclusion}
In this work, one important metric for the unpredictability of a \gls{puf}, the Bit-Alias, is enhanced by introducing a corresponding confidence interval.
A hypothesis test is defined to verify whether each position in the response string complies with given limits on Bit-Alias or entropy.
The suggested confidence interval is applied to previous work, demonstrating that even the most elaborate large-scale tests only reach an accuracy of approximately $\pm 0.1$ for their estimations of Bit-Alias at $\alpha = 0.01$.
These results, together with the proposed hypothesis test, emphasize the demand for a high number of test devices in \gls{puf} research.


\section*{Acknowledgment}
{\small This work was partly funded by the German Research Foundation (DFG) through grant number SI~2064/1-1.
Permanent \textsc{id} and revision date of this document:\\
001624aa04a823c02fd54d3ca29883b932c6f9ee

\isodate\printdate{2019-04-16}}



\bibliographystyle{IEEEtran}
\bibliography{IEEEabrv,abbreviateAuthors,docear_filtered,paper}
%
%
%

\end{document}